# Evaluating e-Government Services in Kurdistan Institution for Strategic Studies and Scientific Research Using the EGOVSAT Model

Aram M. Ahmed[1], Bryar A. Hassan[2], Dr. Soran A. Saeed[3], and Awin A. Saeed[4]

[1]Kurdistan Institution for Strategic Studies and Scientific Research, Department of Information Technology,
Building No. 10, Alley 60, Gullabax 335, Shorsh St., Opposite Shoresh Hospital
Sulaimani, Kurdistan Region, Iraq
*aramahmed@kissr.edu.krd*

[2]Kurdistan Institution for Strategic Studies and Scientific Research, Department of Information Technology,
Building No. 10, Alley 60, Gullabax 335, Shorsh St., Opposite Shoresh Hospital
Sulaimani, Kurdistan Region, Iraq
*bryar.hassan@kissr.edu.krd*

[3]Sulaimani Polytechnic University, Technical College of Engineering,
Qirga,Wrme St. , Mardin 327 , Alley 76
*soran.saeed@spu.edu.iq*

[4]Gorannet Company for Internet and Broadband Services,
Salim Street, Faruq Holding, Flat No. 2B
*aween.saeed@gorannet.net*

**Abstract:** *Office automation is an initiative used to digitally deliver services to citizens, private and public sectors. It is used to digitally collect, store, create, and manipulate office information as a need of accomplishing basic tasks. Azya Office Automation has been implemented as a pilot project in Kurdistan Institution for Strategic Studies and Scientific Research (KISSR) since 2013. The efficiency of governance in Kurdistan Institution for Strategic Studies and Scientific Research has been improved, thanks to its implementation. The aims of this research paper is to evaluate user satisfaction of this software and identify its significant predictors using EGOVSAT Model. The user satisfaction of this model encompasses five main parts, which are utility, reliability, efficiency, customization, and flexibility. For that purpose, a detailed survey is conducted to measure the level of user satisfaction. A total of sixteen questions have distributed among forty one users of the software in KISSR. In order to evaluate the software, three measurement have been used which are reliability test, regression analysis and correlation analysis. The results indicate that the software is successful to a decent extent based on user satisfaction feedbacks obtained by using EGOVSAT Model.*

## 1. INTRODUCTION

E-Government is a strategy to improve the quality, efficiency, transparency and effectiveness in their services. Specifically, Office Automation is able to facilitate towards the achievement of the economic and social development goals, business and civil society at local, state, national and international levels. In Kurdistan Region of Iraq, the initiative has to some extend increase the efficiency of the government in providing better services to its citizens. Nonetheless, low uptake versus high investment deemed the introduction of e-Government a strategy that has yet to be optimized. Meanwhile, Kurdistan Region has a strategy for automating the public sectors so as to improve the efficiency, effectiveness, quality, and transparency in their services [1]. Though the both Iraqi and Kurdistan Region governments have increased efforts for using and adopting e-Government services among employees, the adoption rate might remain low. Barriers and perceived benefits may influence the adoption rate which include factors such as content, design, social status, language, differences in competence and skills, differences in network access, differences in motivation, and other factors. However, research studies have shown that a strategic way to increase participation is by placing the user needs at the center of the development of electronic public services. The level of user satisfaction is an important indicator to further usage and adoption on a large scale basis. Thus, this research paper evaluates and identify and evaluate the key factors in influencing employee (user) satisfaction towards e-Government services in Kurdistan Institution for Strategic Studies and Scientific Research (KISSR) by using the EGOVSAT model.

## 2. AZYA OFFICE AUTOMATION

Azya Office Automation is a government-to-government (G2G) service provided by a local company used to digitally store, create, manipulate, and collect office information needed in order to accomplish basic tasks [2]. It gives services to the organizations of a government in a way that it fastens the workflow of the administrative and managerial processes. In general, Azya Office Automation has three roles. Firstly, Azya Office Automation can accomplish the managerial and administrative tasks in Kurdistan Institution faster than before. Secondly, the need for a large staff can be eliminated in Kurdistan Institution. Thirdly, it needs less storage to store data. Lastly, multiple employees of Kurdistan Institution can simultaneously edit and update data. A short description features of Azya Office Automation is shown in Table 1.

Table 1: Features of Azya Office Automation

| Feature | Value |
| --- | --- |
| Multiple platform | MS Windows |
| Multi-browser | Internet Explorer |
| Word Processor Support | Yes |
| Multi-device accessibility | PC |
| Form generator | Yes |
| Report generator | Yes |
| language support | Kurdish |
| Integration with any solution | Yes |
| Database | MS. SQL Server |
| Programming language | ASP.Net |
| Web based support | Yes |

Furthermore, Azya Office Automation is a web-based initiative used by only the employees of Kurdistan Institution to run the managerial and administrative need of Kurdistan Institution. In that case, the employees act as the citizens. Figure 1 shows the main web based interface of Azya Office Automaiton.

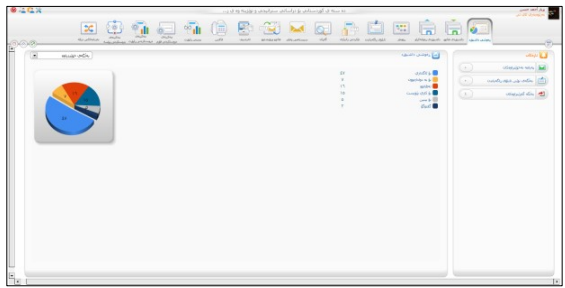

Figure 1 Main interface of Azya Office Automation [3]

Technically, Azya Office Automation works work as an online web based application. A client-server network architecture is used as a communication model to connect the clients to the system, it supports infinite number of users and groups with allocation of user roles and privileges. Figure 2 shows a general technical view of Azya Office Automation.

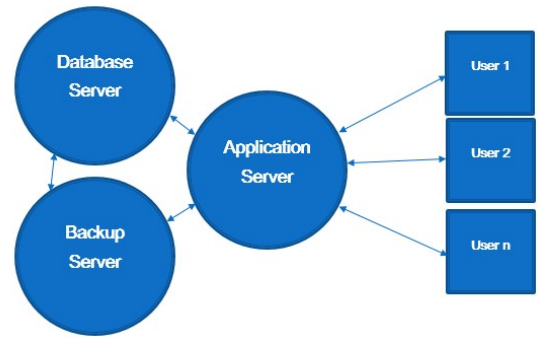

Figure 2 Technical view of Azya Office Automation

Managerially, Azya Office Automation works as focal application where users are connected to each other and can use the system based on the privilege that they have. It has import and export systems to be connected with the other systems. It also has arching system. Refer to figure 3.

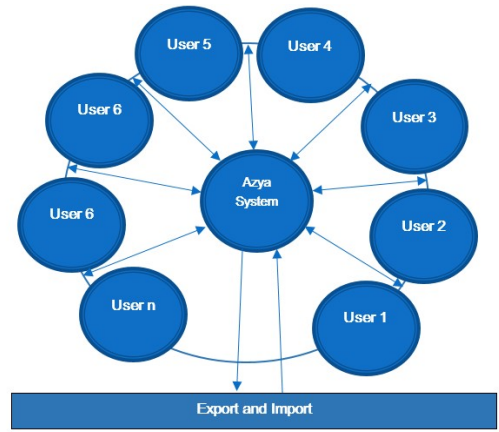

Figure 3 Managerial view of Azya Office Automation

### 3. KURDISTAN INSTITUTION FOR STRATEGIC STUDIES AND SCIENTIFIC RESEARCH (KISSR)

Kurdistan Institution for Strategic Studies and Scientific Research (KISSR) is one of the Ministry of Higher Education and Scientific Research - KRG's public institution located in the city of Sulaymaniyah in Kurdistan Region – Iraq [4]. It is one of the important scientific and cultural institution in Kurdistan region as well as Iraq which is specialized for scientific research and strategy studies. The aim of Kurdistan Institution for Strategic Studies and Scientific Research (KISSR) are to conduct researches in the fields of pure sciences, engineering, medical sciences, and strategic studies.

According to the national development report of Kurdistan Regional Government [1], the vision of implementing e-government and e-organization is focusing to improve the services of public sector effectively and efficiently. Kurdistan Institution is the first governmental organization in Kurdistan Region and Iraq that has used Azya Office

Automation since the late of 2013 to automate managerial and administrative processes. This implies that Azya Office Automation is an applicable system in e-government system.

## 4. RELATED WORK

E-Government adoption is affected by the elements that provide perceived ease of use and perceived usefulness, by user characteristics such as sense of perceived risk, feeling of perceived control, and also by user satisfaction with the quality of its services [12]. E-Government adoption requires that users show higher levels of satisfaction with the online service provided by the government. Also, a higher level of user satisfaction will increase the rate of e-Government adoption [4]. In the previous research works, various techniques, such as Technology Acceptance Model (TAM), McLean and Delone Model, SERVQUAL and EGOVSAT Model, or an integrated model are used to measure user satisfaction [6,7,13,14]. The measurement of user satisfaction can have a meaningful and objective feedback on user's expectation and preference. Furthermore, e-Government performance will be evaluated in relation to set of satisfaction dimensions that indicate the strong and the weak factors affecting user satisfaction of e-Government service [8]. Because the EGOVSAT model has performance dimensions (constructs) [6], it is proposed for evaluating the performance of a web-based e-government system with a citizen-centric or user-centric approach to focus on the process and the outcome of the interaction [4,5].

## 5. RESEARCH MODEL

The EGOVSAT is a model proposed for evaluating the performance of a web-based e-government system with a citizen-centric or user-centric approach to focus on the process as well as the outcome of the interaction [4,5]. According to [6], the EGOVSAT model has five performance dimensions (constructs) for user satisfaction, which are utility, efficiency, reliability, flexibility, and customization. Also, each contract includes some emotional dimensions (items). The total number of items for all the five constructs are sixteen items. In addition, each construct is used to examine a specific feature of e-Government service. Utility construct is used to examine whether the web-based Office Automation is usable or not. Reliability construct is to examine whether the web-based Office Automation functions well or not with regard to technology and the content accuracy. The efficiency is to examine the organization and accessibility of the features and information available in the web-based Office Automation. For the customization construct, it is to examine the customizability offered to the users in both information content and access methods. Flexibility construct refers to whether the web-based office automation provides choice of state a need and provides dynamic information. The model is presented in Figure 4.

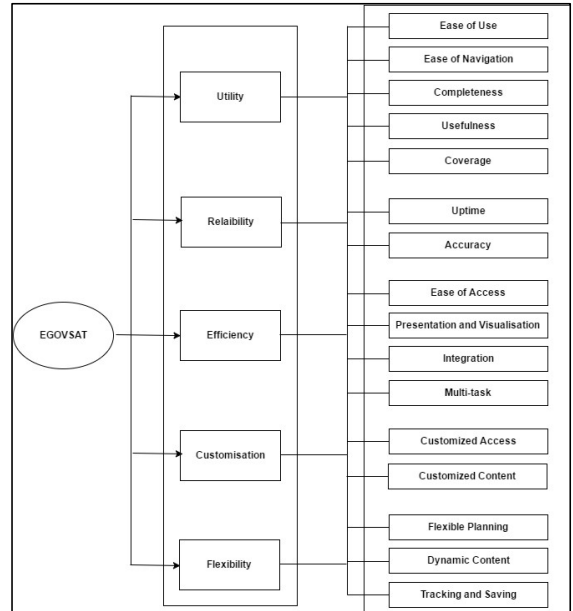

**Figure 4** The EGOVSAT model

Incidentally, high level of user satisfaction for the services of e-Government is an important key to measure the adoption of the services provided by government. A higher level of user satisfaction will increase the rate of e-Government adoption [7]. Meanwhile, user satisfaction is considered as a performance measurement of e-Government services and can be measured by various techniques. For examples, McLean and Delone Model, Technology Acceptance Model (TAM), SERVQUAL, and EGOVSAT Model can be used to measure user satisfaction. Furthermore, by measuring user satisfaction it will have an immediate, meaningful and objective feedback about user's preference and expectation. On the other hand, e-Government performance will be evaluated in relation to set of satisfaction dimensions that indicate the strong and the weak factors affecting user satisfaction of e-Government service [8]. Thus, The EGOVSAT model is used in this study with the aim of providing a scale by which government to citizen web based initiatives can be evaluated in terms of satisfaction derived by users (employees) [4].

## 6. RESEARCH METHODOLOGY

In this research paper, a survey is employed to collect data from the users of Azya Office Automation.

### a. Data Collection

In this study, questionnaires have distributed to all the users of Azya Office Automation in Kurdistan Institution for Strategic Studies and Scientific Research. This questionnaire based on the constructs and their items of the EGOVSAT model. A relevant data were collected via the questionnaire and analyzed using Statistical Software Package for Social Sciences (SPSS).

b. Methodology

In the constructs, each item is measured on ten points (10 = strongly agree to 1 = strongly disagree)

### 7. DATA ANALYSIS

The data is analyzed using statistical methods, users' profile data, reliability analysis, correlation analysis, and multiple regression analysis.

a. User Data Profile

The users' profile of Azya Office Automation of the surveyed respondents is presented in Table 2. The total number of the users is forty one users.

The gender distribution is 53.66% male and 46.34% female. Further, 53.66% are from scientific education background; whereas 46.34% are from humanities education background. The age of users are varied. 7.32% are 22-29 years old; 53.66% are 30-39 years old; 24.39% are 40-49 years old; 12.2% are 50-59 years old; and 2.44% are above 60 years old. Their educational level is also varied from Post doctorate to Baccalaureate degrees. At the same time, each of them have different background in computer skills. More than half of them have intermediate level in using computer. Table 2 presents the user profiles of Azya Office Automation.

Table 2: User Data Profile

|  |  | Frequency | Percent |
|---|---|---|---|
| Education | Science | 22 | 53.66 |
|  | Humanities | 19 | 46.34 |
| Gender | Male | 22 | 53.66 |
|  | Female | 19 | 46.34 |
| Age | 22-29 | 3 | 7.32 |
|  | 30-39 | 22 | 53.66 |
|  | 40-49 | 10 | 24.39 |
|  | 50-59 | 5 | 12.2 |
|  | Above 60 | 1 | 2.44 |
| Education level | Post doctorate | 1 | 2.44 |
|  | Doctorate | 4 | 9.76 |
|  | Master | 6 | 14.63 |
|  | Higher diploma | 0 | 0 |
|  | Bachelor | 18 | 43.9 |
|  | Diploma | 4 | 12.2 |
|  | Baccalaureate | 7 | 17.07 |
| Computer skills | None | 0 | 0 |
|  | Basic | 1 | 2.44 |
|  | Intermediate | 23 | 56.1 |
|  | Advance | 9 | 21.95 |
|  | Expert | 8 | 19.51 |

b. Reliability Analysis

Cronbach's alpha is a measure of internal consistency, that is, how closely related a set of items are as a group [15]. It also is as a coefficient of reliability or consistency to be a measure of scale reliability. Cronbach's alpha can be written as a function of the number of test items and the average inter-correlation among the items. Equation (1) shows the formula of standardized Cronbach's alpha.

$$\alpha = \frac{N.\bar{c}}{v+(N-1).\bar{c}} \quad (1)$$

Here N is equal to the number of items, c-bar is the average inter-item covariance among the items and v-bar equals the average variance.

In the equation (1), if the number of items are increased, Cronbach's alpha will be increased. Additionally, if the average inter-item correlation is low, alpha will be low. As the average inter-item correlation increases, Cronbach's alpha increases as well (holding the number of items constant).

In this study, Cronbach's alpha can be used to evaluate the reliability of the items [9]. A Cronbach's score which is considered as reliable is 0.7 or higher. Table 3 shows the result of reliability statistics for all the five constructs. The score ranges is between 0.925 and 0.669. The Utility contract score the highest, whereas the Flexibility construct score the lowest. The test of reliability analysis predicts that only three constructs are considered as reliable as well as suitable for further analysis, which are Utility, Efficiency, and Customization. Nonetheless, Reliability and Flexibility constructs are marginally reliable. The detailed result is shown in Table 3.

Table 3: Result for reliability analysis

| Construct | No. of Items | Crohbach's Alpha |
|---|---|---|
| Utility | 5 | 0.925 |
| Reliability | 2 | 0.695 |
| Efficiency | 4 | 0.848 |
| Customization | 2 | 0.895 |
| Flexibility | 3 | 0.669 |

c. Correlation Analysis

Pearson's Correlation Coefficient (r), the square root of r2, is a measure of association between two interval-ratio variables. Equation (2) represents the way of calculating correlation analysis between two items, such as X and Y [16].

$$r = \frac{\text{cov}(X,Y)}{s.d.(X)*s.d.(Y)} \quad (2)$$

According the equation (1) of correlation analysis, the measure is symmetrical and there is no specification of independent or dependent variables. Additionally, the range of r is from −1.0 to +1.0. The sign (±) indicates direction. The closer the number is to ±1.0 the stronger the association between X and Y.

In this study, correlation analysis is basically used to determine the strength as well as the direction between the constructs. The purpose of correlation analysis is to test hypotheses of the EGOVSAT model. This is conducted by finding the correlation ratio between the predicator variables. As discussed previously, the sixteen questions are the predictor variables. (The questions are named as Q11 until Q53). The result of correlation analysis is shown in table 4. Where each predictor variable is compared with all the other predictor variables and the correlation ratio between them are found accordingly. For example, the correlation between a predictor variable and itself is ONE because they are fully correlated and so on.

Table 4: Result of correlation analysis

|     | Q11 | Q12 | Q13 | Q14 | Q15 | Q21 | Q22 | Q31 | Q32 | Q33 | Q34 | Q41 | Q42 | Q51 | Q52 | Q53 |
|-----|-----|-----|-----|-----|-----|-----|-----|-----|-----|-----|-----|-----|-----|-----|-----|-----|
| Q11 | 1 | | | | | | | | | | | | | | | |
| Q12 | .766 | 1 | | | | | | | | | | | | | | |
| Q13 | .635 | .759 | 1 | | | | | | | | | | | | | |
| Q14 | .578 | .699 | .675 | 1 | | | | | | | | | | | | |
| Q15 | .754 | .749 | .659 | .892 | 1 | | | | | | | | | | | |
| Q21 | .446 | .163 | .176 | .115 | .272 | 1 | | | | | | | | | | |
| Q22 | .611 | .663 | .532 | .553 | .662 | .532 | 1 | | | | | | | | | |
| Q31 | .368 | .517 | .507 | .433 | .362 | .171 | .313 | 1 | | | | | | | | |
| Q32 | .338 | .590 | .554 | .411 | .426 | .387 | .536 | .540 | 1 | | | | | | | |
| Q33 | .435 | .614 | .649 | .423 | .435 | .240 | .498 | .616 | .816 | 1 | | | | | | |
| Q34 | .559 | .625 | .493 | .385 | .532 | .366 | .610 | .482 | .706 | .771 | 1 | | | | | |
| Q41 | .542 | .749 | .656 | .566 | .570 | .035 | .436 | .653 | .673 | .713 | .676 | 1 | | | | |
| Q42 | .425 | .579 | .481 | .425 | .405 | -.131 | .268 | .624 | .391 | .516 | .475 | .819 | 1 | | | |
| Q51 | .149 | .362 | .193 | .074 | .066 | -.047 | .187 | .499 | .407 | .480 | .428 | .534 | .528 | 1 | | |
| Q52 | .291 | .357 | .283 | .109 | .186 | .096 | .255 | .274 | .226 | .247 | .159 | .409 | .483 | .421 | 1 | |
| Q53 | .486 | .541 | .492 | .304 | .459 | .124 | .578 | .201 | .560 | .630 | .593 | .604 | .464 | .435 | .367 | 1 |

### d. Multiple Regression Analysis

Ordinary least squares linear regression is the most widely used type of regression for predicting the value of one dependent variable from the value of one independent variable. It is also widely used for predicting the value of one dependent variable from the values of two or more independent variables. When there are two or more independent variables, it is called multiple regression.

The elements of multiple regression equation is shown in the equation (3) [17].

$$Y = a + b_1 X_1 + b_2 X_2 + b_3 X_3 \qquad (3)$$

Where:
- Y is the value of the Dependent variable (Y), what is being predicted or explained
- a (Alpha) is the Constant or intercept
- b1 is the Slope (Beta coefficient) for X1
- X1 First independent variable that is explaining the variance in Y
- b2 is the Slope (Beta coefficient) for X2
- X2 Second independent variable that is explaining the variance in Y
- b3 is the Slope (Beta coefficient) for X3
- X3 Third independent variable that is explaining the variance in Y
- s.e.b1 standard error of coefficient b1
- s.e.b2 standard error of coefficient b2
- s.e.b3 standard error of coefficient b3
- R2 The proportion of the variance in the values of the dependent variable (Y) explained by all the independent variables (Xs) in the equation together
- F Whether the equation as a whole is statistically significant in explaining Y

For further investigation of the relationship between the user satisfactions (EGOVSAT) and the constructs (utility, reliability, efficiency, customization, and flexibility). These constructs are represented as independent variables, whereas the EGOVSAT is represented as dependent variables. The EGOVSAT is measured by taking the means of system quality, overall satisfaction, and information quality as well as overall satisfaction.

A Cronbach Alpha score of 0.695 was obtained which indicates the accuracy of the measuring instruments for this construct multi co-linearity because there was no considerable correlations (R>.9) among the variables as shown in table 5.

Table 5: Regression on EGOVSAT

| Model | | Unstandardized Coefficients | | Standardized Coefficients | t | Sig. |
|---|---|---|---|---|---|---|
| | | B | Std. Error | Beta | | |
| 1 | (Constant) | 5.245 | .728 | | 7.199 | .000 |
| | Utility | .420 | .119 | .576 | 3.563 | .001 |
| | Efficiency | .398 | .138 | .525 | 2.878 | .007 |
| | Customization | -.544 | .152 | -.724 | -3.581 | .001 |

a. Dependent Variable: Reliability
R squared=0.509, Adjusted R squared=.454,
F Value=9.329**, N=40
***Significant at  p=.000

## 8. RESULT

In the previous section, three statistical approaches which are reliability analysis, correlation analysis, and multiple regression analysis have been applied to evaluate the questionnaire.

Firstly, as it is clearly shown in table 6, the test of Crohbach's Alpha shows that only three constructs which are Utility, Efficiency, and Customization, have scored more than the threshold (0.7) and are considered to be reliable. However, Reliability and Flexibility constructs fell below the threshold (0.7) and are considered to be marginally reliable.

Table 6: Result of Crohbach's Alpha

| Construct | Result & Threshold Comparison | Result of Crohbach's Alpha |
|---|---|---|
| Utility | 0.925 > 0.7 | Fully reliable |
| Reliability | 0.695 < 0.7 | Marginally reliable |
| Efficiency | 0.848 > 0.7 | Fully reliable |
| Customization | 0.895 > 0.7 | Fully reliable |
| Flexibility | 0.669 < 0.7 | Marginally reliable |

Secondly, the aim of applying correlation analysis is to test hypotheses of the EGOVSAT model. Figure 5 shows that all the predictor factors in relation with the user satisfaction of Azya Office Automation are significant and positive. That means, all of the predictor variables are correlated to each other to some extent. The closer predictor variable to ±1.0, the stronger the association between the predictor variables. As it is clearly show in graph X, the predictors are very close to each other in the range of 0 to 0.8 in X axis and -0.2 to 1 in Y axis.

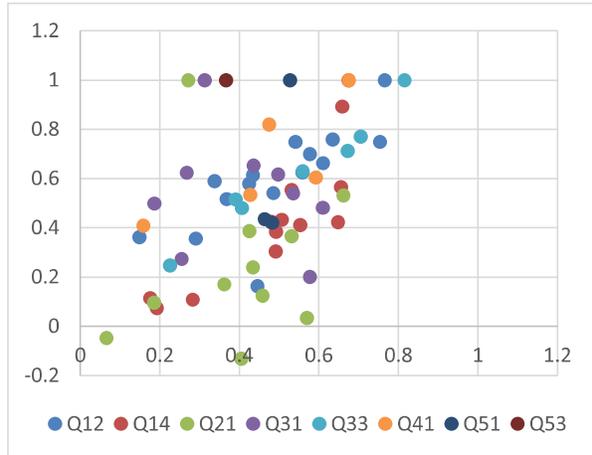

**Figure 4** Correlation Analysis Result

Thirdly, the linear combination of factors was significantly related to user satisfaction on e-Government services accounting for approximately 48 percent of the variance. All the predictors are significant at the 95 percent confidence level and exhibit a positive relationship with EGOVSAT. Nonetheless, predictors that will have a higher magnitude in bringing about change (standard deviations) in EGOVSAT are utility, flexibility, customization and efficiency consecutively.

## 9. DISCUSSION

Even though the result is given based on the measurements of EGOVSAT model for user satisfaction, the result might be varied or the same if McLean and Delone Model, Technology Acceptance Model (TAM), or SERVQUAL is applied. More than one model can be a better practice to be applied to achieve more accurate result. Moreover, the number of users used in the survey is forty one users. More users can be taken if they exist. Also, the type of survey is just question and answer with a giving score from one to ten by the users. Alternatively, the system can be evaluated by giving some specific tasks to do them in a specific time and consequently score their results.

The data of the survey is analyzed using three statistical approaches, which are reliability analysis, correlation analysis, and multiple regression analysis. This can be analyzed using some more statistical approaches to get more accurate results. Depending on the score giving importance of predictors may change according to that user's usage , but from this model we obtained that reliability , Utility , Efficiency , Customization and Flexibility in EGOVSAT model improvements have a great importance to user satisfaction. As an alternative of these models, an automated tool can be applied to evaluate such kind of e-Government services. Nevertheless, there is not any automated tool to test and validate all of the constructs of EGOVSAT model.

## 10. CONCLUSION

In this study, the EGOVSAT model is used in this study with the aim of providing a scale by which government to citizen web based initiatives can be evaluated in terms of satisfaction derived by users. As the source of data, a survey is used to collect data from the users of Azya Office Automation based on the constructs and their items of the EGOVSAT model. Then, the data is analyzed using statistical methods, users' profile data, reliability analysis, correlation analysis, and multiple regression analysis.

Thus, user satisfaction is one of the vital indicator to give a general idea as to how well the government has transformed its services according to its needs. The level of satisfaction will also be an important indicator to further usage and adoption on a large scale basis. The EGOVSAT Model presented in this paper has identified significant predictors in influencing employee (user) satisfaction.

## ACKNOWLEDGEMENT

This study has been done in Kurdistan Institution for Strategic Studies and Scientific Research (KISSR) and Sulaimani Polytechnic University (SPU). The authors wish to express their deep thanks to Professor Dr. Polla Khanaqa, the head of the KISSR; Assist. Professor Dr. Alan F. Ali, the president of SPU for their kind support in conducting this study.

**AUTHORS**

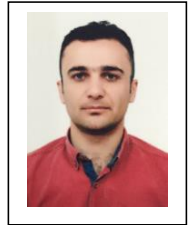

**Aram M. Ahmed** received his BSc in Software Engineering from Koya university/Iraq in 2010; MSc in Computer Network Technology at Northumbria University/United Kingdom in 2013. Currently, he is a PhD candidate in Computer Science at KISSR and Sulaimani Polytechnic University.

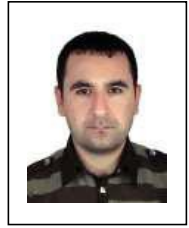

**Bryar Hassan** received MSc Software Engineering in 2013 from the University of Southampton, UK. He is currently a lecturer/researcher and PhD candidate in Kurdistan Institution for Strategic Studies and Scientific Research as we as the director of Information Technology department in Kurdistan Institution. His research interests are in Semantic Web, Web 3.0, Information Systems, and E-Government.

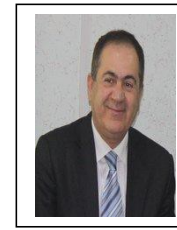

**Dr. Soran Saeed** received the PhD in Computer Science from University of Greenwich University - London, UK in 2006. He is assistant professor of Computer Science and his current position is the Vice President for Scientific Affairs and Higher Education at Sulaimani Polytechnic University (SPU). He is also head the board of e-Court System at Sulaimani court developing by AKTORS Company from Estonia.

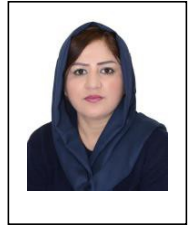

**Awin A. Saeed** MSc candidate in Information Technology from Sulaimani Polytechnic University and Kurdistan Institution for Strategic Studies and Scientific Research. She works as a visiting assistant researcher at Computer Science institute, Sulaimani Polytechnic University (SPU) as well as Technical Support from Gorannet Company for Internet and Broadband Services. Her research interests are Computer Networks, Information System, and E-Government Services.